\documentclass[oneside,a4paper,11pt,shownumbers,manualsort]{article}
\usepackage{latexsym}
\usepackage{euscript}
\usepackage{epsfig,amsmath,amssymb}
\usepackage{slashed}
 
\date{\today}
\renewcommand\({\left(}
\renewcommand\){\right)}
\renewcommand\[{\left[}
\renewcommand\]{\right]}

\newcommand{\dd}{{\rm d}}
\newcommand{\e}{{\rm e}}
\def\be{\begin{equation}}
\def\ee{\end{equation}}
\def\bea{\begin{eqnarray}}
\def\eea{\end{eqnarray}}

\newcommand\vp{\varphi}

\newcommand{\eref}[1]{(\ref{#1})}

\newcommand\mcL{\mathcal L}

\newcommand\mcO{\mathcal O}

\long\def\symbolfootnote[#1]#2{\begingroup%
\def\thefootnote{\fnsymbol{footnote}}\footnote[#1]{#2}\endgroup} 

 \large\normalsize

\begin{document}

\begin{center}

{\Large \bf Zero modes on cosmic string loops}

\vspace*{7mm} {Marieke Postma $^{a}$
\symbolfootnote[1]{{E-mail:mpostma@nikhef.nl}}
and Betti Hartmann $^{b}$}
\symbolfootnote[2]{E-mail:betti.hartmann@lmpt.univ-tours.fr}
\vspace*{.25cm}

${}^{a)}${\it NIKHEF, Kruislaan 409, NL-1098 SJ Amsterdam, The Netherlands}\\
\vspace*{.1cm} 
${}^{b)}${\it Laboratoire de Math\'ematiques et Physique Th\'eorique,
Universit\'e de Tours, Parc Grandmont, 37200 Tours, France}

\end{center}

\begin{abstract}
We study the spectrum of fermionic modes on cosmic string loops.  We
find no fermionic zero modes nor massive bound states --- this implies
that vortons stabilized by fermionic currents do not exist. We have
also studied kink-(anti)kink and vortex-(anti)vortex systems and find
that all systems that have vanishing net topological charge do not
support fermionic bound modes.
\end{abstract}


\section{Introduction}

Cosmic string loops have raised a lot of interest in the past decades.
They are believed to form in the evolution of cosmic string networks.
While standard loops of cosmic string eventually decay, it has been
suggested that fermionic or bosonic currents ``living'' on the loop of
string can stabilize it against decay. This was first put forward in
\cite{davis} within a model of superconducting strings \cite{witten}.
Such stable cosmic string loops are called {\it vortons} (see
\cite{VS} and references therein).  In the case of bosonic
currents, a scalar field condensate builds up in the core of the
string and leads to a conserved (Noether) current. Vortons have been
constructed explicitly in a scalar field model with $U(1)\times U(1)$
symmetry \cite{ls}.
 
Fermionic currents appear due to the coupling of the string forming
Higgs and/or gauge field to fermions.  The aim of this paper is the
construction of the spectrum of both massless and massive fermionic
bound states on cosmic string loops.  For an infinitely long straight
string the spectrum is well known: the number of zero modes is given
by an index theorem and equals the winding number of the
string~\cite{rossi,Weinberg}, and there are massive bound states as
well~\cite{stephen,ringeval}.  The equivalent question for the string
loop is hardly studied.  An incomplete analysis of this case was done
in \cite{Gadde,pr}; they found that the number of zero modes is equal
to the local winding number of the string.  Although this seems to
agree with the result as for the infinitely long straight string, we
note that the straight string has a net winding whereas the loop does
not.  In \cite{pr} the existence of fermionic zero modes for a curved
string was established analyzing small curvature corrections to the
straight string solution.  However, it is not clear whether this
analysis can be extended to string loops.  In this paper, we
asymptotically solve the fermionic equations of motion in the presence
of a string loop.  In contrast to \cite{Gadde}, we will show that no
fermionic zero modes or massive bounds states exists on cosmic string
loops.  And thus no fermionic vortons exists.

In fact, we would not expect the existence of fermionic zero modes on
string loops.  The number of zero modes is an adiabatic invariant.
One can adiabatically change the string loop to the trivial vacuum
background, and the latter supports no bound states. Saying it in
other words, the topological charge of the whole system is zero, and
it follows from an index theorem \cite{Weinberg} that there are no
bound states. This statement can be verified explicitly in the much
simpler case of a kink-antikink system.  This system carries no net
topological charge, it can be adiabatically deformed to the trivial
vacuum. And indeed, solving the equations of motion, we find that the
kink-antikink does not support fermionic bound states. The
kink-antikink system is quite analogue to the set-up with two parallel
straight strings with opposite winding number. For these simpler
systems, we find that there are no fermionic zero modes for systems
with vanishing net topological charge.

Our paper is organized as follows. In section 2 we discuss the
spectrum of fermion bound states on kinks and antikinks.  In section 3
we move on to cosmic strings, and discuss single straight strings, two
parallel strings as well as string loops. We end with conclusions in
section 4.

\section{Kinks and Antikinks}

We first examine the relatively simple system of kinks and
antikinks. We expect that many of the results, which as we will see
are topological in nature, carry over to the case of cosmic strings to
be discussed in the next section.

\subsection{Bosonic background}

Consider a bosonic background of the form
\be
\phi = \eta \Big[ \tanh(a(z-z_1)) 
+ B \bigg( \tanh\(a(z+z_1)\) +1\Big) \bigg].
\label{phidw}
\ee
with $\eta >0$. $B=0$ corresponds to the well known kink solution of
the $\phi^4$-theory located at $z=z_1$; the antikink solution is
obtained by replacing $\eta \to -\eta$.  $B=-1$ describes a
kink-antikink configuration.  This is only a solution of the field
equations in the limit $z_1 \to \infty$. Finally, $B=1$ corresponds to
a kink-kink solution.  This set-up needs a bosonic potential with at
least three minima, and thus does not occur in the $\phi^4$-theory.
In principle, the bosonic potential can be reconstructed from the
Bogomolny equations via $\partial_z \phi = \sqrt{2V}$.  

The ``thin wall'' approximation, i.e. the limit that the domain wall
is infinitely thin, is equivalent to the limit $a \to \infty$.
Eq. \eref{phidw} then reduces to
\be
\phi = \left\{
\begin{array}{lc}
- \eta & \qquad z < -z_1, \\
\eta & \qquad -z_1 < z < z_1, \\
\eta (1+2B) & \qquad z_1 < z .
\end{array}
\label{tophat}
\right.
\ee

\subsection{Zero mode solutions}
\label{s:kink_zm}

Consider a fermion with a Yukawa interaction $\mcL \ni h \phi
\bar{\psi} \psi$.  The Dirac equation in the background of kinks and
antikinks described by (\ref{phidw}) reads
\be
\[ i \gamma^\mu \partial_\mu - h \phi \] \psi =0.
\ee
To find the zero mode solutions we separate the longitudinal and
transverse coordinates, and write
\be
\psi = \alpha(t,x,y) \beta(z) \xi,
\ee
with $\xi$ a constant spinor.  Here $\alpha(t,x,y)$ solves the
longitudinal Dirac equation and gives the dispersion relation.  To
find the zero mode solution we set $E=0$, $k_L=0$ for the moment (so
that it is trivially solved), and $\alpha =1$; the case $E\neq 0$ is
discussed in the next subsection.  To solve the transverse part of the
Dirac equations we introduce the eigenspinors
\be
\gamma^z \xi_\pm = \pm i \xi_\pm
\label{gammaz}  .
\ee
The two projection eigenstates decouple in the Dirac equation, which
becomes
$\partial_z \beta_\pm = \mp h \phi \beta_\pm$.
It has as solution
\be
\beta_\pm(z) = N \exp(\mp B m_\psi z) \cosh\[a(z-z_1)\]^{\mp B m_\psi/a}
\cosh\[a(z+z_1)\]^{\mp m_\psi/a},
\label{betaB}
\ee
with $N$ a constant normalization constant, and $m_\psi = h \eta$ the
 vacuum fermion mass.

In the chiral basis for $\gamma$ matrices, see appendix \ref{s:conv},
the eigenspinors of the projection operator (\ref{gammaz}) are of the
form
\be
\xi^1_+ = \( \begin{matrix} 0\\ i\\ 0\\ 1 \end{matrix} \), \quad
\xi^2_+ = \( \begin{matrix} -i\\ 0\\ 1\\ 0 \end{matrix}\), \quad
\xi^1_- = \( \begin{matrix} 0\\ -i\\ 0\\ 1 \end{matrix} \), \quad
\xi^2_- = \( \begin{matrix} i\\ 0\\ 1\\ 0 \end{matrix}\).
\label{dwspinors}
\ee
There are four independent solutions $\psi_\pm^i = \beta_\pm
\xi_\pm^i$ with $i=1,2$.  They are normalizable if
\be \int_{-\infty}^\infty \dd z \, \psi_m^\dagger \psi_m = 
\int_{-\infty}^\infty \dd z \, |\beta_\pm|^2 \delta_{mn} < \infty
\ee
with $\psi_m$ labeling the 4 states $\psi^i_\pm$.

\paragraph{Kink or antikink.}
To get the kink zero-mode solution we set $B=0$ in \eref{betaB} to
obtain
\be
\beta_\pm = N \cosh\[a(z+z_1)\]^{\mp m_\psi/a}.
\label{betakink}
\ee
The asymptotic behavior of the solutions is $\beta_\pm \propto
\exp(\mp m_\psi |z|)$. The $\beta_-$ solution is non-normalizable,
while $\beta_+$ corresponds to a zero mode solution localized at the
kink at $z=-z_1$. Two independent real solutions, or one complex
solution, exist, constructed from $\psi_+^i = \beta_+ \xi_+^i$ with
$i=1,2$.  To find the solution for an antikink, we replace $\eta \to
-\eta$. Now $\beta_-$ is the localized normalizable solution, while
$\beta_+$ is non-normalizable. The profiles of the fermionic zero
modes for kinks and antikinks respectively are shown in Fig.~1.

\begin{figure}
\includegraphics[width=8cm]{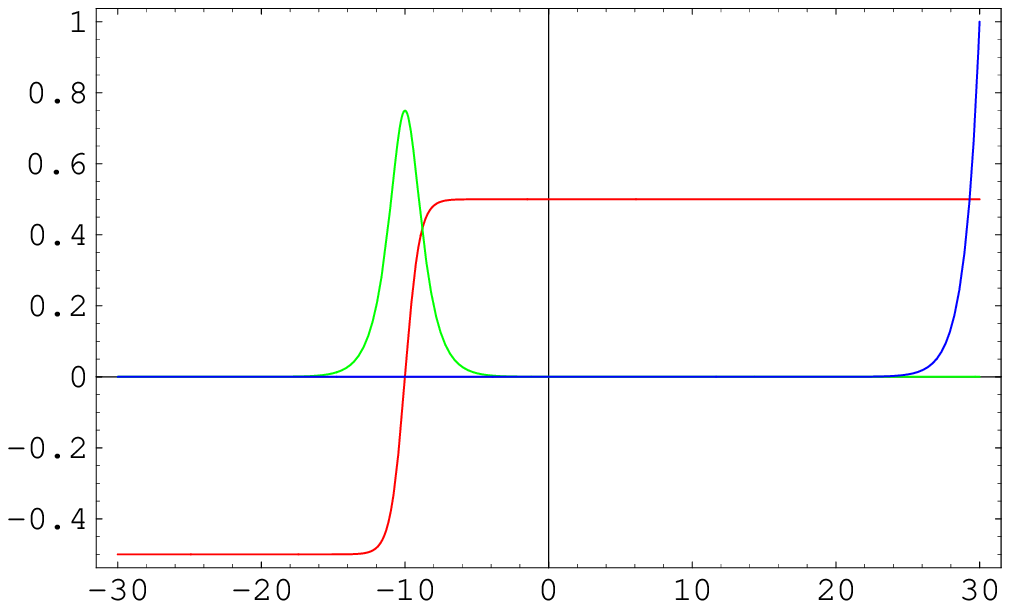} 
\includegraphics[width=8cm]{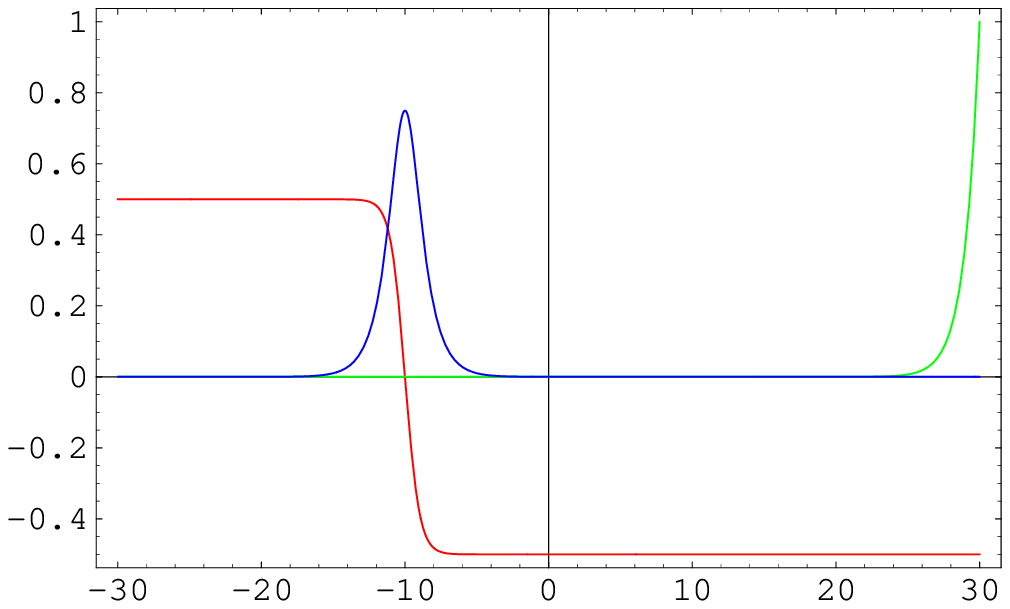}
\\
\includegraphics[width=8cm]{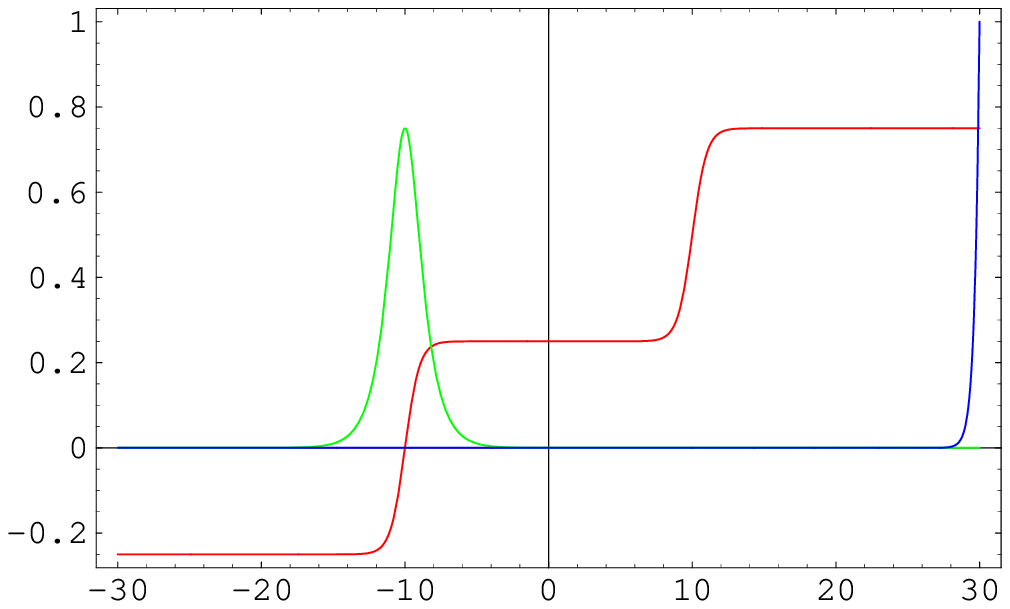} 
\includegraphics[width=8cm]{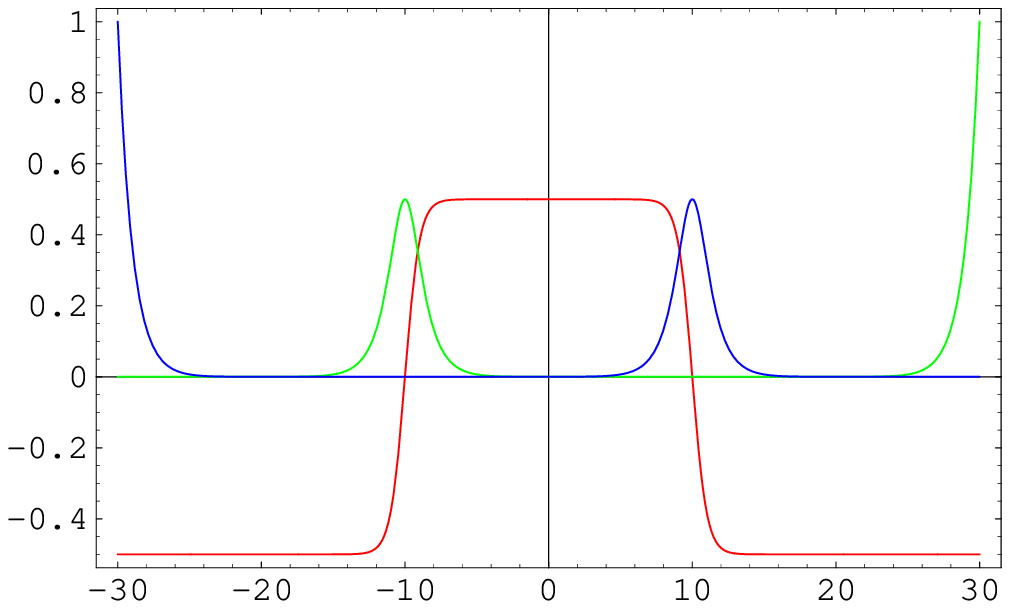}
\caption{The profiles of the fermionic zero mode solutions $\beta_+$
(green) and $\beta_-$ (blue) are shown together with the bosonic
background $\phi$ (red) for the kink (top left), antikink (top right),
kink-kink (bottom left) and kink-antikink (bottom right)
configurations, respectively.  We have chosen $z_1=10$,
$\eta=h=a=1$.}
\end{figure}

\paragraph{Kink-kink.}
The kink-kink configuration corresponds to $B=1$.  The asymptotic
behavior of the solutions \eref{betaB} then reads $\beta_-
\propto\exp(m_\psi |z|)$ which is non-normalizable, and $\beta_+
\propto \exp(-3 m_\psi |z|)$ which is normalizable.  Both the bosonic
background and the fermionic solutions are invariant under $z_1
\leftrightarrow -z_1$.  This implies that we find only one complex
zero mode solution, which is localized around $z=-z_1$. The profile of
the zero mode solution is shown in Fig.~1.

Consider now the limit $z_1 \to 0$.  The bosonic background then
corresponds to a kink-configuration with $\phi$ interpolating between
$-\eta$ and $3\eta$. The zero mode solution becomes $\beta_\pm =
\exp(\mp m_\psi z) \cosh[az]^{\mp 2m_\psi/a}$.  As expected $\beta_+$
has the usual form of a normalizable zero mode on a kink
\eref{betakink}. The factor 2 difference in the exponent of the
$\cosh$ is due to the fact that the topological charge $Q \propto
\phi(\infty) - \phi(-\infty)$ is twice as big as for a usual kink.
The factor $\exp(\mp m_\psi x)$ comes from the fact that the $B=1$
double kink solution is shifted horizontally by $\eta$ compared to the
$B=0$ kink background, it is of the form $\phi = \eta + 2
\eta \tanh(az)$.

\paragraph{Kink-antikink.}
The kink-antikink configuration corresponds to $B=-1$. Both zero mode
solutions, although they are peaked at the kink or antikink, are
non-normalizable since $\beta_\pm(\pm \infty) \propto \exp(m_\psi
|z|)$.  Thus no normalizable zero mode solution exists.

This result can be easily understood. The zero mode solution on a kink
is a $\xi_{+}$-spinor, on an antikink a $\xi_{-}$-spinor.  So pasting
the kink and antikink together, one can never match the zero mode
solutions at the origin because of the orthogonal spinors.  This
problem is independent of the distance between the kinks, in
particular it does not disappear in the limit $z_{1} \to \infty$.  And
whereas the kink zero mode solution falls off exponentially at
infinity $\beta_+ \propto \e^{- h \phi z} $ with $\phi = \eta$, this
no longer holds for the kink-antikink system. Indeed, extending the
wave function past the antikink, $\phi$ flips sign, and $\beta_+
\propto \e^{+ h \eta z}$ blows up.

Maybe in a cosmological setting the zero mode solution does not have
to be extended all the way to infinity as there is a natural cutoff.
The cutoff can for example be due to the presence of yet another kink
or anti-kink (cf. the argument for the existence of a network of
global strings), or (in a higher dimensional setting) the boundary of
spacetime.  We can compare the height of the wave function localized
at the kink at $z = -z_1$ with how it blows up at infinity if it is
extended past the antikink at $z = z_1$. This gives
\be
\frac{\beta_+ (z = -z_1)}{\beta_+ (z \to \infty)}
 = \frac{ \e^{m_\psi z_1}} {\e^{m_\psi (2z -z_1)}},
\ee
where we took the thin wall limit $a \to \infty$. It follows that
$|\psi(-z_1)|= |\psi(z\to\infty)|$ for $z = 3 z_1$. Thus the cutoff
has to be smaller $z_{\rm cutoff} < 3 z_1 $ for the fermion to be
localized on the kink at $z = -z_1$.

\subsection{Massive bound states}

We now proceed to consider non-zero energy solutions. In the rest
frame $\vec{k}=0$, and the Dirac equation reads
\be
\[i\gamma^0\partial_0 + i\gamma^z\partial_z -h\phi\] \psi =0.
\ee
Now $\xi^i_\pm$, which are eigenspinors of $\gamma^z$ (see
(\ref{gammaz}), (\ref{dwspinors})), are not eigenspinors of
$\gamma^0$.  The fact that the energy is non-zero mixes the
spinors $\xi_\pm$.  

We write the energy eigenstates as
\be
\psi_E^k =\e^{-i E t}
\Big( \beta^k_+(z) \xi^k_- -i \beta^k_-(z) \xi^k_- \Big)
\ee
with $k=1,2$. The factor $-i$ is put in for future convenience.
$\psi_E^1$ and $\psi_E^2$ decouple in the Dirac equation.  Their
equations of motion are identical.  In the following we will suppress
the superscript, keeping in mind that the equations apply to both.
The energy eigenstates satisfy $\hat H \psi_E = E \psi_E$ with
\be
\hat H  = i \partial_t = \gamma^0 (-i\gamma^z \partial_z + h \phi) .
\ee
This gives the following set of coupled equations
\bea
E \beta_+ &=& (\partial_z - h \phi) \beta_- \nonumber \\
-E \beta_- &=& (\partial_z + h \phi) \beta_+
\label{eom1}
\eea
where we used $\gamma^0 \xi_\pm = \pm i \xi_\mp$.

Let us analyze the asymptotic solutions. In the regions at spatial
infinity, and for the case of multiple (anti)kink systems in between
the (anti)kinks, the background field approaches a constant
$\partial_z \phi =0$.  In this limit the equations of motion
decouple. Eq. \eref{eom1} can be written in the form
\be
\partial^2_z \beta_\pm = (m_f^2 - E^2) \beta_\pm
\ee
which has the solution
\be
\tilde \beta_\pm = U_\pm \e^{\sqrt{m_f^2 - E^2} z} 
+ D_\pm  \e^{-\sqrt{m_f^2 - E^2} z}
\label{UDsol}
\ee
with $m_f = h |\phi|$ the vacuum fermion mass, and $U_\pm, D_\pm$
integration constants. At spatial infinity one of the solutions is
normalizable.  Plugging back in \eref{eom1} we find 
\be
\frac{U_-}{U_+} =  \frac{- \sqrt{m_f^2-E^2} - h \phi}{E},
\qquad
\frac{D_-}{D_+} = \frac{ \sqrt{m_f^2-E^2} - h \phi}{E}.
\label{UDratio}
\ee
In the core of a kink located at $z=0$ the bosonic background is
$\lim_{z\to 0} \phi(0) = \eta a z + {\mathcal O}(z^2)$.  To find the
asymptotic solutions we write $\beta_\pm = b_\pm z^{n_\pm}$ near the
origin, with $n_\pm \geq 0$ to assure regularity at the core.  The
equations of motion (\ref{eom1}) reduce to
\bea
E b_+ z^{n_+} &=& n_- b_- z^{n_- -1} -m_f  a  b_- z^{n_- + 1} ,
\nonumber \\
- E b_- z^{n_-} &=& n_+ b_+ z^{n_+ -1} + m_f a  b_+ z^{n_+ +1} .
\label{dwcore}
\eea
The term proportional to $m_f$ is higher order. There is no general
solution that solves the equations.  There are however two special
solutions.  The first is the zero mode solution \eref{betakink} with
$E = 0, \, b_- =0, n_+ =0$, which sets all lowest order terms to zero.
The second special solution is a bound state solutions with $E \neq 0$
and $n_- =0, \, n_+ = 1$.  This choice allows the lowest order terms
in the first equation in \eref{dwcore} to cancel, whereas in the
second equation the higher order terms come in at lowest order as
well.

\paragraph{Kink or antikink.}

The kink is known to have one bound state solution, which can be
constructed explicitly.  For simplicity consider a kink located at
$z_1 =0$. To find the bound state solution we substitute $\beta_+ = C
\phi \beta_-$ in the Dirac equations \eref{eom1}.  Provided we choose
$C = - E/(a \eta)$ and $E^2 = a\eta(2h - a/\eta)$, the equations are
degenerate and read 
\be
\partial_z \beta_- = \eta \tanh(a z) \( \frac{a}{\eta} -h\) \beta_-
\ee
which has as solution $\beta_-(z) = N \cosh(a z)^{1-h/a}$ with $N$ a
normalization constant. This bound state only exists for $h >
a/(2\eta)$.  There is no bound state solution in the thin wall limit
$a/\eta \to \infty$.  Unlike the zero mode solution, the number of
bound states is not an adiabatic invariant, it is not determined by an
index theorem.

The above result agrees with the asymptotic analysis.  Indeed,
consider a kink located at $z = -z_1$ in the thin wall approximation
(see (\ref{tophat}) with $B=0$).  In region I $z\in [-\infty:0]$ the
bosonic background is $\phi = -\eta$, and in region II $z\in
[0:\infty]$ it is $\phi = \eta$.  The normalizable solution in region
I is $\psi_E^I = \e^{\sqrt{m_f^2 -E^2}z} (U^I_+ \xi_+ +U^I_- \xi_-)$,
the solution in region II is $\psi_E^{II} = \e^{-\sqrt{m_f^2 -E^2}z}
(D^{II}_+ \xi_+ +D^{II}_- \xi_-)$.  Matching the solutions at $z=0$
requires
\be
U_-^I/U_+^I = D_-^{II}/D_+^{II}
\ee
which is impossible for $E \neq 0$.  Setting $E=0$ in the equations of
motion, we see that the first equation of \eref{eom1} gives back the
normalizable zero mode solution. Hence, in the thin wall approximation
we find back the zero mode solution, but do not find additional
massive bound states.

\paragraph{Kink-kink}
Consider now the kink-kink system. Following the same strategy as for
the kink system we write $\beta_+ = C \phi \beta_- $.  Now, however,
we do not find a special solution to the equations of motion
\eref{eom1}.  The difference with the kink case is the appearance of
cross terms $\tanh(a(z-z_1)) \tanh(a(z+z_1))$ which destroy the
solution. One may argue that our ansatz for $\beta_+$ should be
changed for the kink-kink system, but it is fixed by the requirement
that at $z=-z_1$ the solution should approach the kink solution.

In the limit $z_1 \to 0$ the kink-kink system reduces to a single
kink, which has a massive bound state.  This bound state is very
fragile.  It dissapears in the thin wall limit.  And from the
kink-kink analysis it follows that it dissapears as well if the kink
is deformed.

The kink-kink system has one zero mode and no bound states in the
spectrum.

We have also tried to construct solutions to the equations (\ref{eom1})
numerically, but could not find solutions.

\paragraph{Kink-antikink}

The story for the kink-antikink system is similar to the kink-kink
system.  We choose an Ansatz $\beta_+ = C \phi \beta_- $, which is
dictated by the requirement that the solution approaches the kink
solution at the kink at $z = -z_c$.  No bound state solution is found,
compared to the kink case it is the appearance of cross terms
$\tanh(a(z-z_1)) \tanh(a(z+z_1))$ that destroy the solution.

The same conclusion also follows in the thin wall approximation
(\ref{tophat}) with $B=-1$.  The solutions in the three distinct
regions are given by (\ref{UDsol}). The boundary conditions
constrain the solution in the asymptotic regions: $D_\pm =0$ in the $z
< -z_1$ region, and $U_\pm =0$ in the $z > z_1$ region.  Matching the
solutions at both $z=-z_1$ and $z=z_1$, we find that the coefficients
appearing in (\ref{UDsol}) in the middle region $-z_1 < z < z_1$
satisfy $D_+/D_- = U_+/U_- =1$, which is impossible to satisfy, even
for $E=0$.  Hence, there are no bound states at all.

This can be understood as follows. The decaying solution at minus
infinity is $\propto \xi_+$, whereas the solution at plus infinity is
$\propto \xi_-$, and these orthogonal spinors cannot be matched in the
middle. This spinor structure of the decaying solution is determined
by the value $\phi(\pm \infty)$, hence is directly related to
topological arguments. The decaying solution at plus and minus
infinity is the same spinor iff the topological charge is non-zero: $Q
\propto \phi(\infty) - \phi(-\infty) \neq 0 $. For the kink-antikink
system, on the other hand $\phi(\infty) - \phi(-\infty) =0$.  The
kink-antikink system can be adiabatically deformed to the trivial
vacuum.  As the vacuum carries no bound states, the kink anti-kink
system has none either.

The Dirac equation describes a fermion with a spatially varying mass
term, i.e. a fermion living in an effective potential $V_{\rm eff} =
|h\phi|^{2}$.  For the kink-antikink system the effective potential is
a double well potential.  Experience with the analogue quantum
mechanical double well systems suggests there should be two bound
states, split in energy with the energy difference going to zero in
the limit that the two wells are taken infinitely far apart. It seems
from our results above that this intuition fails for the fermion zero
modes. Why?

The analogue with the quantum double well system breaks down.  The
double well potential in quantum mechanical problems can be
adiabatically transformed by reducing the distance between the two
wells, and for $z_{1} = 0$ one ends up with a single well.  The number
of bound states is preserved.  What happens is that the two lowest
bound states of the single well become the two energy split bound
states of the double wells.  Now taking $z_{1} \to 0$ in the
kink-antikink system, they will annihilate, and one ends up with the
trivial vacuum. The vacuum does not support bound states, so it is not
surprising that there are no zero modes on the kink-antikink system.
The number of zero modes is given by an index theorem, it is
$(1/\eta)(\phi(\infty) -\phi(-\infty))$, which is zero for a
kink-antikink configuration.

Note that our results disagree with those stated in
\cite{jackiw,altschul,choi}. In \cite{altschul} the energy is
calculated for $\Psi = (\psi_+ \pm \psi_-)_{B=0}$ states, which is
split in energy and goes to zero in the $z_1 \to \infty$ limit, just
as in a double well potential.  However, these are zero mode solutions
to a background with only one kink or antikink, they do not satisfy
the Dirac equation for the kink-antikink system.  Using instead $\Psi
= (\psi_+ \pm \psi_-)_{B=-1}$, the wave function is non-renormalizable
and the energy blows up.

\section{Cosmic strings}

Having built up intuition on the existence of fermionic modes on
relatively simple domain wall system, let us see whether the same kind
of arguments apply to systems with cosmic strings.  

Consider two parallel strings, one with winding number (= topological
charge) $+n$ and one with winding number $-n$.  There is no net
winding around the string, and thus --- by analogy with domain walls
--- we do not expect fermionic zero modes in this system.  The system
can be deformed adiabatically to the trivial vacuum which does not
support bound states.  Looking at the spinor structure, zero modes on
a string with $n>0$ ($n<0$), have positive (negative) chirality.
Pasting a string and ``antistring'' together, the zero modes cannot be
matched in the middle because of the orthogonal spinor structure ---
in close analogy with the domain wall.  This problem is not
ameliorated in the limit that the strings are taken far apart. The
solution that is localized at the string does not resolve the angular
dependency at the antistring, and vice versa.  It might be that a
bound state with non-zero energy can cure this problem since it mixes
the two chiralities.  But we have seen that for the kink-antikink
system this was not the case.


The system of two parallel strings is closely related to the loop of
cosmic string.  In both cases there is no net winding number
separating the system from the vacuum. Hence, we expect that the
spectrum of fermionic bound states found on two anti-parallel strings
(none!)  is the same as for the string loop. This implies that there
are no fermionic vortons, no loops of cosmic strings stabilized by
fermionic currents.\\

In this section we study fermionic bound states on straight strings
and loops of string.  In the next subsection, we discuss the bosonic
string background.  In section \ref{s:zmstraight} and \ref{s:zmloop}
we derive the fermionic spectrum for straight strings and a string
loop respectively. To count the number of bound states an asymptotic
analysis suffices. A numerical analysis is needed to obtain the
profile of the bound states solutions, which can be done in the full
system including back reaction effects.  This involves systems of
nonlinear ordinary and partial differential equations for straight
strings and string loops, respectively; it is left as a future
project.

\subsection{Bosonic string background}

In this section we discuss the asymptotic behavior of local cosmic
string, both for straight strings and for loops. We neglect gravity
effects, as well as back reaction effects from fermions (which will not
effect the number of bound states). Moreover, we restrict ourselves to
strings satisfying the Bogomolny limit. Note that in this limit there
is no force between parallel straight strings.  Details on the
cylindrical and toroidal coordinates used can be found in Appendix
\ref{s:coord}.

The flat space-time metric is
\be
\dd s^2 = \dd t^2 - h_1^2 \dd x_1^2 - h_2^2 \dd x_2^2 - h_3^2 \dd x_1^2.
\label{3metric}
\ee
The Lagrangian describing the gauged string (Nielsen-Olesen string)
reads
\be
\mcL = (D_\mu \phi)(D^\mu \phi)^* - \frac14 F_{\mu \nu}F^{\mu \nu} 
- \frac{\lambda^2}{2} \( |\phi|^2 - \eta^2 \)^2,
\ee
with $D_\mu \phi = (\partial_\mu + ie A_\mu) \phi$ and $F_{\mu \nu} =
\partial_\mu A_\nu -\partial_\nu A_\mu$.  The static cosmic string
configuration aligned with the 3-axis, has a magnetic field confined
to the string core and aligned with the 3-axis $B \equiv B_3 =
F_{12}/(h_1 h_2)$.  The Higgs field winds around the $x_3$ axis with
winding number $n$ (we take $n>0$ without loss of generality). The
string energy per unit length is
\bea
\mu &=& 
\int \sqrt{-g_{_T}} \dd x_1 \dd x_2 \left\{ 
 (D_\mu \phi)(D^\mu \phi)^*  - \frac14 F_{\mu \nu}F^{\mu \nu} 
+ \frac{\lambda^2}{2} \( |\phi|^2 - \eta^2 \)^2 \right\}
\nonumber \\
&=& 
\int \sqrt{-g_{_T}} \dd x_1 \dd x_2 \left \{ 
 \left|\frac{D_1 \phi}{h_1} \right|^2 
+ \left| \frac{D_2 \phi}{h_2} \right|^2  
+ \frac12 \left| \frac{F_{12}} {h_1 h_2} \right|^2 
+ \frac{\lambda^2}{2} \( |\phi|^2 - \eta^2 \)^2 \right\}
\eea
where the integration is over the transverse coordinates $x_1,x_2$
which span the plane perpendicular to the string. The metric on this
transverse plane satisfies $\sqrt{-g_{_T}} = h_1 h_2$.  Following the
arguments of Bogomolny --- using partial integration and the identity
$[D_1,D_2] \phi = i e F_{12} \phi$ --- we write the tension as
\bea
\mu &=& \int  \dd x_1 \dd x_2 \, h_1 h_2 \Bigg \{
\left| \( \frac{D_1}{h_1} + i \frac{D_2}{h_2} \) \phi \right |^2 +
\frac12 \left| \frac{F_{12}}{h_1 h_2} - e ( |\phi|^2 - \eta^2) \right|^2
 \nonumber \\
&&+ \frac{\lambda^2 -e^2}{2} ( |\phi|^2 - \eta^2)^2
-e \eta^2 \frac{F_{12}}{h_1 h_2}
\Bigg\} .
\eea
We have omitted the boundary term.  The last term in the above
expression is
\be
Q = -e  \eta^2 \int  \dd x_1 \dd x_2  {F_{12}} 
= - e \eta^2 \int B = 2 \pi n \eta^2
\ee
for a string with winding number $n$. Note that this is a topological
number for a straight string, but not for a loop of string.  For a
loop, the string goes up and down through the transverse plane, and
these contributions cancel. Hence, a string loop can decay, there is
no topological barrier separating it from the vacuum.  In the BPS
limit $e =\lambda$, the third term cancels, and it follows that $\mu
\geq Q$.  The minimum energy configuration with $\mu = Q$ satisfies
the BPS equations
\bea
\( \frac{D_1}{h_1} + i \frac{D_2}{h_2} \) \phi &=& 0
\\
\frac{F_{12}}{h_1 h_2} - e ( |\phi|^2 - \eta^2) &=& 0
\eea

\subsubsection{Straight string aligned with $z$-axis}

We use cylindrical coordinates with $x_1=r$, $x_2=\vp$ and
$x_3=z$. The results for this case are well known, however, we repeat
them here for completeness.  The scale factors are $h_r=h_z =1$ and
$h_\vp = r$.  Using the standard string Ansatz
\bea
\phi &=& \eta f(r) \e^{i n \vp} \\
A_\vp &=& - \frac{n}{e} a(r)
\label{fa_straight}
\eea
the BPS equations become
\bea
(\partial_r  + \frac{i}{r} D_\vp) \phi &=& 0
\qquad \Rightarrow \qquad  
\partial_r f - \frac{n}{r}(1-a) f = 0
\nonumber \\
\frac{\partial_r A_\vp}{r} - e (|\phi|^2 - \eta^2) &=& 0
\qquad \Rightarrow \qquad  
n\frac{\partial_r a}{r} - e^2 \eta^2 (1-f^2) = 0
\eea
with boundary conditions $f(0)=a(0) =0$ and $f(\infty)=a(\infty)
=1$. Note that the energy of the configurations remains finite,
because the gauge field cancels the kinetic energy of the Higgs field
(due to the phase dependence) at infinity.  The magnetic field is
given by $B = \nabla \times A = \partial_r A_\vp$.  The solution is
symmetric under rotations around the $z$-axis.

First, we consider the equations in the limit $r\to 0$.  The term
proportional to $a$ in the first equation tends to zero, and it
follows $f \propto r^{|n|}$.  The second term in the second equation
becomes $r$-independent and can cancel the first term for $a \propto
r^2$, i.e.
\be ({\rm for} \;r\to 0 ) \;\;\; \left\{
\begin{array}{ll}
f \propto r^{|n|} \\
a \propto r^2
\end{array}
\right.  
\label{straightA}
\ee
Far away from the string core the fields tend to their vacuum values
$f\to 1$, $a \to 1$.  To see how the fields approach these values we
substitute $f = 1 - c_1 e^{- c_3 r}$ and $a= 1- r c_2 \e^{-c_3 r}$
into the BPS equations. Upon taking the limit $r \to \infty$ we find
$c_3 = \sqrt{2} e \eta = m_{\!A}$ the gauge boson mass (in the BPS
limit $m_{\!A} = m_\phi$) and $c_2/c_1 = m_{\!A}/n$.  We have:
\be
({\rm for} \; r\to \infty)   \;\;\;
\left\{
\begin{array}{ll}
f - 1 \propto \e^{-m_{\!A} r}\\
a - 1   \propto (m_{\!A} r) \e^{-m_{\!A} r}
\end{array}
\right.
\ee

\subsubsection{Cosmic string loop}

Consider a loop of cosmic string lying in the $x$-$y$-plane. We
introduce toroidal coordinates with $x_1=v$, $x_2=u$, $x_3=\vp$ such
that the symmetry axis of the loop is the $z$-axis and the loop is
aligned with the $\vp$-direction. This set up is certainly less
symmetric than that of the straight string since the fields now depend
on the angle $u$ winding around the string (in fact, to find the
cosmic string loop, one has to solve partial differential equations in
contrast to ordinary differential equations for the straight case). If
the radius of the loop becomes very large, the $u$-dependence will
practically disappear.  The string Ansatz can be generalized to
\bea
\phi &=& \eta f(u,v) \e^{i n u} \\
A_u &=& - \frac{n}{e} a(u,v)
\label{fa_loop}
\eea
and we used the ``radial'' gauge $A_v=0$.  The boundary conditions are
\begin{eqnarray}
\lim_{v\to \infty}  f,a &=& 0 \ \ \ \ {\rm string} \ {\rm core}, \nonumber \\
\lim_{(v,u) \to (0,0)} f,a &=& 1 \ \ \ \ {\rm spatial} \ {\rm infinity}, 
\nonumber \\
\lim_{(v,u) \to (0,\pi)} f,a &=& 1 \ \ \ \ {\rm origin}.
\end{eqnarray}
Note that we have assumed here that the loop radius is much larger
than the string width $\kappa \gg m_A^{-1}$, so that at the origin
(and along the full $z$-axis) the bosonic fields approach their vacuum
values.  This requirement can be relaxed, all that is needed is that
the fields are regular at the origin.  However, in that case it is
doubtful how well a cosmic string loop pictures the situation.

The gauge field $A_u$ is non-zero to cancel the kinetic term of the
Higgs field $\phi$ at spatial infinity. The magnetic field is $B_\vp =
- \partial_v A_u(u,v)$. The BPS equations are
\bea
\partial_v f + n(1-a) f =0,  \qquad \partial_u f = 0
\label{fv}
\\
n \partial_v a+ h_u h_v e^2 \eta^2 (1-f^2) =0
\eea
We get two equations for $f$ from splitting the real and imaginary
parts. It follows that the Higgs profile function is $u$-independent;
the only $u$ dependence of the Higgs is in the winding phase. The
Higgs profile is independent of the scale factors $h_i$, and thus of
the loop radius $\kappa$. The gauge profile function $a$ is
generically $u$-dependent since $h_u h_v$ depends on $u$. At infinity
$h_u h_v$ is only a function of $v$ and both $f$ and $a$ are functions
of $v$ only.  Note that the $u$-independence of $a$ far away from the
string loop reflects the fact that there is no net winding around the
string loop, no net topological charge.

Let us consider the asymptotic solutions. For details of the behavior
of the toroidal coordinates in these limits see Appendix \ref{s:coord}.
First consider the limit $v\to \infty$ in which the string core is
approached and $f\to 0$, $a \to 0$. The BPS equations reduce to
\be
\partial_v f+nf = 0, \qquad 
\partial_v a + {\rm const.}\cdot \e^{-2v}= 0 
\ee
which has as solution
\be
f \propto e^{-nv} \sim \(\frac{\delta r}{\kappa}\)^n ,
\qquad
a \propto e^{-2v} \sim \(\frac{\delta r}{\kappa}\)^2
\ee
with $\delta r = |r -\kappa|$ denoting the distance from the string
core. The behavior is the same as in the core of a straight string
\eref{straightA}.  The fields approach their vacuum values far away
from the string loop, corresponding to the limit $v\rightarrow 0$,
$u\rightarrow 0$, and can be written in the form $f = 1 - \tilde{f}$,
$a = 1 - \tilde{a}/n$.  The BPS equations become
\be
\partial_v\tilde{f}=\tilde{a},
\qquad
\partial_v\tilde{a} = h_u h_v m_{\!A}^2 \tilde{f}
\ee
with $m_{\!A} = \sqrt{2} e\eta$.  At spatial infinity, corresponding
to $u=v\to0$, the factor $h_u h_v$ tends to $(\kappa/v^2)^2$.  The
equations are then solved to get (up to a normalization constant)
\bea
f-1 &\propto&  n \e^{- m_{\!A} \kappa/v} \sim  n \e^{- m_{\!A} r}
\nonumber \\
a -1 &\propto&  \frac{m_{\!A}\kappa}{ v^{2}} 
\e^{- m_{\!A} a/v} 
\sim \frac{ m_{\!A}r^2}{\kappa}   \e^{- m_{\!A} r}
\eea
At the origin, corresponding to $u\to \pi, v\to 0$, the factor $h_u
h_v$ tends to $(\kappa/2)^2$.  The solutions are up to a
renormalization constant
\bea
f-1 &\propto& \sinh\(\frac{m_{\!A} \kappa v}{2}\)
\sim  \sinh(m_{\!A} r) \sim (m_{\!A}r)
\nonumber \\
a -1 &\propto&  \(\frac{m_{\!A}\kappa}2\) 
\( \cosh\(\frac{m_{\!A}\kappa v}2\) -1 \) \sim
\(\frac{m_{\!A}\kappa}2\) 
\( \cosh(m_{\!A} r) - 1 \) \sim \(\frac{m_{\!A}\kappa}2\)  (m_{\!A} r)^2
\eea
The approach to the vacuum is exponentially fast at spatial infinity
(just as it is in the case of a straight string), but slower --- power
law --- at the origin.  The solutions are valid for $v \ll 1$ which
corresponds to $r \gg \kappa$ at spatial infinity, and $r \ll \kappa$
at the origin. The width of the string is $r \sim m_{\!A}^{-1}$.



\subsection{Fermionic spectrum}

Consider a Dirac spinor $\psi = {\psi_L \choose \psi_R}$, with
$\psi_L,\psi_R$ left and right handed two-component Weyl spinors.  The
fermion is charged under the $U(1)$ of the string. The Lagrangian reads
\be
\mcL = \psi_L^\dagger \bar{\sigma} \cdot D \psi_L +
\psi_R^\dagger \sigma \cdot D \psi_R
-\lambda \phi \psi_L^\dagger \psi_R 
- \lambda \phi^* \psi_R^\dagger \psi_L
\ee
with $D_\mu \psi_{L,R} = (\partial_\mu + iq_{L,R}eA_\mu)\psi_{L,R}$.
Gauge invariance requires $1-q_L + q_R = 0$, where we have normalized
the Higgs charge to unity. For a Dirac spinor the two Weyl fermions
$\psi_L,\psi_R$ are independent, and can have different charges. For a
Majorana fermion $q_L = -q_R =1/2$ and $\psi_R = i\sigma^2 \psi_L^*$
are charged conjugate of each other, so that the Majorana spinor
satisfies the reality condition $\psi_M= \psi_M^c$.  The reality
condition decreases the degrees of freedom.  The equations of motion
are
\bea
i \bar{\sigma} \cdot D \psi_L - \lambda \phi \psi_R &=& 0
\nonumber \\
i \sigma \cdot D \psi_R - (\lambda \phi)^* \psi_L &=& 0
\label{diraceq}
\eea
For a Majorana the two equations are not independent, but complex
conjugate of each other.

\subsubsection{Single straight string}
\label{s:zmstraight}

The transverse and longitudinal part of the solution decouple, and we
can write
\be
\psi_{L} = \alpha(z,t) \beta_{L}(r,\vp) \xi_{L},
\qquad
\psi_{R} = \alpha(z,t) \beta_{R} (r,\vp) \xi_{R}.
\label{decomp}
\ee
Here $\alpha$ and $\beta_{L,R}$ are scalar functions of the transverse
and longitudinal coordinates respectively. The constant Weyl spinors
$\xi_{L,R}$ are eigenvectors of the projection operator
\be
\sigma^0 \sigma^z \xi_L = \pm \xi_L,  
\qquad
\sigma^0 \sigma^z \xi_R = \mp \xi_R,  
\label{projection}
\ee
with the upper sign for a vortex ($n>0$) and lower sign for an
anti-vortex ($n<0$).  Explicitly $\xi_L = {1 \choose 0}$ and $\xi_R =
{0 \choose -i}$ for $n>0$, and $\xi_L = {0 \choose 1}$ and $\xi_R =
{-i \choose 0}$ for $n<0$, where the factor $i$ in the right-handed
spinors is for convenience.  The Weyl spinors have the following
properties: $i \sigma^0 \sigma^1 \sigma^2 \sigma^3 \xi_{L,R} =
-\xi_{L,R}$ (chirality eigenstates), and $\sigma^r \xi_L = i \e^{\pm i
\vp} \xi_R$, $\sigma^r \xi_R = -i \e^{\mp i \vp} \xi_L$, $\sigma^{\vp}
\xi_L= \mp e^{\pm i\vp}\xi_R$ and $\sigma^{\vp} \xi_R= \mp e^{\mp
i\vp}\xi_L$. Using this the transverse part of the Dirac equation can
be written as
\bea
\( \partial_r \pm i \frac{D_\vp}{h_\vp} \) \beta_L
&=& \lambda \eta f(r) \e^{\pm i(|n|-1)\vp} \beta_R
\nonumber \\
\( \partial_r \mp i \frac{D_\vp}{h_\vp} \) \beta_R
&=& \lambda \eta f(r) \e^{\mp i(|n|-1)\vp} \beta_L
\label{Dstring}
\eea
with $f(r)$ the profile function of the Higgs field in
\eref{fa_straight}, and as before the upper (lower) sign is for a
vortex (anti-vortex).  The phase-dependence can be resolved by
choosing
\be
\beta_{L,R}(r,\vp) = b_{L,R}(r) \e^{il_{L,R}\vp},
\qquad
l_L = l_R \pm (|n|-1).
\ee
The equations reduce to
\bea
\( \partial_r \mp \frac{(l_L - n q_L a) }{r} \) b_L 
&=& m_f f b_R
\nonumber\\
\( \partial_r \pm \frac{(l_R - n q_R a) }{r} \) b_R 
&=& m_f f b_L
\eea
where we used $h_\vp = r$, and $m_f = \lambda \eta$ is the vacuum
fermion mass.

We can solve the Dirac equation asymptotically.  For simplicity we
restrict the analysis to the vortex solution with $n>0$.  At $r \to
\infty$ the functions $f\to 1$, $a\to 1$ and the $1/r$ term in the
equations of motion is subdominant.  The solution is
\be
b_{L,R} = \e^{\pm m_f r}
\ee
There is one renormalizable solution.  At the origin $f \sim r^n$ and
$a\sim r^2$.  We write $b_{L,R} \sim r^{c_{L,R}}$. Then the equations
of motion reduce to 
\bea
c_L r^{c_L-1} - l_L r^{c_L-1} &=& m_f r^{c_R+n}
\nonumber \\
c_R r^{c_R-1} + l_R r^{c_R-1} &=& m_f r^{c_L+n}
\eea
A singular solutions exist for $c_L = l_L$, $c_R = -l_R$.
Two regular solutions exist for $c_L = l_L$, $c_R = l_L + n +1 $ and
$c_R = - l_R = n-1-l_L$, $c_L =l_L+2n$.  They connect to the one
normalizable solution at infinity. Both need to be normalizable at the
origin
\be
\int \sqrt{-g_{_T}} \dd^2 x_{_T} |\psi_{L,R}|^2 < \infty
\qquad \Rightarrow \qquad
\int \dd r\, r |b_{L,R}|^2 < \infty
\qquad \Rightarrow \qquad
2 c_{L,R} + 1 > -1.
\ee
This is satisfied for
\be
-1 < l_L <n.
\ee
Thus a string with winding number $n$ has $|n|$ zero mode solutions
with $l=0,..,|n|-1$. The zero mode spinor is proportional to $\xi_+$
for a vortex with $n>0$, and proportional to $\xi_-$ for an antivortex
with $n<0$.

For $|n|=1$ the zero-mode solution does not have any angular
dependence. The Majorana solution (with $\psi_R = \psi_L^c$) can be
given analytically \cite{rossi}:
\be
\beta_\pm = 
\e^{ \int_0^r \dd r' \( -\frac{a(r')}{2r'}+m_f f(r')\)}.
\ee

The longitudinal part of the Dirac equation $(\sigma^0 \partial_0 -
{\sigma}^z \partial_z)\psi_{L} = 0$ becomes
\be
\( \partial_0 \mp \partial_z\)\alpha = 0
\quad \Longrightarrow \quad
\alpha = \alpha(t\pm z).
\ee
The zero mode on a string (antistring) moves at the speed of light in
the minus (plus) direction along the string and has a dispersion
relation $E = |k|$.

\subsubsection{Two parallel strings}

Consider now two parallel strings, one string located at $x=x_1$
and one string or antistring at $x =-x_1$.  If the distance between
them is much larger than the string width, the bosonic background is
well approximated by the two separated string solutions pasted
together at $x=0$.

\paragraph{String-string.}
Consider two parallel strings which both have winding number $n =
+1$. The bosonic background is then of the form $\phi = f_1 \e^{i \vp}
+ f_2 \e^{i \vp}$ with $f_i \to 0$ at string $i$, and approaching the
vacuum $f_i \to 1$ everywhere else.

The zero mode solutions localized at the string at $x=-x_1$ and the
zero mode at the second string localized at $x=x_1$ have the same
spinor structure. Moreover, near the string the angular dependence is
resolved.  Hence, they can be pasted together at $x=0$. Indeed, we
expect a zero-mode solution of the form $\psi_{R,L} = (\beta_1 \xi_+ +
\beta_2 \xi_+)_{R,L}$, with $\beta_{i}$ localized at string $i$.  The
equations of motion read:
\be
\( \partial_r + i \frac{D_\vp}{h_\vp} \) (\beta_1+\beta_2)_{L}
= \lambda \eta (f_1+f_2) (\beta_1+\beta_2)_{R}
\ee
and a similar equation for the right handed spinors (interchanging $L
\leftrightarrow R$ and a different sign in front of $D_\vp$, see
\eref{Dstring}). The two solutions decouple and are approximately
given by the solutions for the single string; since $f_1 \beta_2 \to
0$ at the location of the first string and $f_2 \beta_1 \to 0$ at the
location of the second, the corrections are small.  The important
thing to note is that the zero mode solution localized at one string
can be extended without problems to the region where the other string
is, as the angular dependence is still resolved there --- this is in
sharp contrast with the string antistring system to be discussed
shortly.

This agrees with topological arguments.  In the limit $x_1 \to 0$ the
system reduces to a single string with winding number $n=2$.  Such a
string has two zero mode solutions.  As the strings are adiabatically
separated, the number of zero modes remains constant. The two zero
modes are the states localized at either one of the strings. This is
different from the kink case. For strings, the number of fermionic
zero modes equals the topological charge which is in $\mathbb{Z}$,
while for kinks the topological charge is in $\mathbb{Z}_2$
(asymptotic vacua either different or the same) and the number of zero
modes does not exceed unity.

\paragraph{String-Antistring.}
Consider now two parallel strings with opposite winding number, one
string with $n = +1$ and the other with $n=-1$.  The zero mode
solution localized on the string and antistring have orthogonal spinor
structure and decouple in the equations of motion. We can thus
consider them separately. We write the zero-mode solution in the form
$\psi_{L,R} = (\beta_1 \xi_+ + \beta_2 \xi_-)_{L,R}$, with $\beta_{i}$
localized at string $i$.  Let's concentrate on the $\beta_1$ solution;
the $\beta_2$ solutions are similar.  The equation of motion for
$\beta_1$ is
\be
\( \partial_r + i \frac{D_\vp}{h_\vp} \) \beta_{1L}
= \lambda \eta (f_1+f_2 \e^{-2i\vp}) \beta_{1R}
\ee
and a similar expression for $L \leftrightarrow R$.  The $\beta_1$
solution does not solve the equations of motion near the antistring,
where $n=-1$ and the angular dependence is unresolved. Although
$\beta_1 \to 0$ near the antistring, it is not equal to zero. Thus,
the zero mode solution disappears from the spectrum.

This agrees with the adiabatic argument. In the limit $x_1 \to 0$ one
ends up with the vacuum ($n=0$). Since there is no topological charge,
there are no zero modes. As the strings are slowly pulled apart the
number of zero modes remains constant.

The adiabatic argument should also hold for bound state solutions with
$E\neq 0$.  The non-zero energy mixes $\xi_+$ and $\xi_-$ spinors,
which in principle might lead to a solution valid in the whole domain.
We thus try the Ansatz
\be 
\psi_{L,R} = \e^{-iE t} \( \beta_1 \xi_+ +  \beta_2 \xi_- \)_{L,R}.
\ee
The equations of motion are then of the form:
\bea
i E \beta_{1L} + (\partial_r - \frac{i}{2r} D_\vp)\e^{-i \vp} \beta_{2L}
&=& m_f (f_1 \e^{i \vp} + f_2 \e^{-i \vp}) \beta_{2R}
\nonumber \\
iE \beta_{2L} + (\partial_r + \frac{i}{2r} D_\vp)\e^{i \vp} \beta_{1L}
&=& m_f (f_1 \e^{i \vp} + f_2 \e^{-i \vp}) \beta_{1R}
\eea
and a similar equation for $L \leftrightarrow R$.  To resolve the
angular dependence we write $\beta_i = \e^{i l_i \vp} b_i$, with $i =
\{1,2,3,4\} =\{1L,2L,1R,2R\}$.  Then the angular structure of the Dirac
equation is of the form
\bea
\e^{i l_1 \vp} + \e^{i(l_2-1)\vp} &=& \e^{i(l_4 +1)\vp} +  \e^{i(l_4 -1)\vp}
\nonumber\\
\e^{i l_2 \vp} + \e^{i(l_1+1)\vp} &=& \e^{i(l_3 +1)\vp} +  \e^{i(l_3 -1)\vp}
\eea
Set $l_1 = l_4 +1$ and $l_2 = l_4$. This resolves the angular
dependence of the first equation above. Inserting this into the second
equation, one finds that no $l_3$ exists that can solve the angular
dependence. Likewise, for $l_1 = l_4 -1$ and $l_2 = l_4 +2$, the
angular dependence for the first equation can be solved, but again no
$l_3$ exists that solves the angular dependence of the second
equation.  There is no choice of $l_i$ that resolves the angular
dependence! And thus we conclude that the string antistring system
does not support massive bound states either.~\footnote{The single
straight string is known to have massive bound states
\cite{stephen,ringeval}.  These states do resolve the angular
dependence in the Dirac equations.}

\subsection{Cosmic string loops}
\label{s:zmloop}

In this section we discuss fermionic bound states on a loop of cosmic
string.  The total winding number of the string loop is zero, the
bosonic configuration can be continuously deformed towards the trivial
vacuum.  And thus we do not expect any bound states to exist, in
direct analogy with the kink antikink and vortex anti-vortex system.

It is clear that the loop does not support zero mode solutions moving
at the speed of light \cite{barr,postma}.  Massless particles which go
at the speed of light move along straight lines, and not along a
curved path.  (Indeed, along a curved path $\dot{k} \neq 0$ and a
dispersion relation $E^2 = k^2$ is not possible). But a priori this
does not exclude a zero mode solution in the limit $k \to 0$.  In
Ref.~\cite{Gadde} the existence of zero mode solutions was claimed.
However, their analysis is incomplete.  They failed to check the
behavior of the zero mode solutions at the center of the string, where
the solutions is singular.  We show this explicitly in the appendix
\ref{s:gadde}, where we give the asymptotic solutions of the Dirac
equation.

The obstruction to bound states is the same as for the vortex
anti-vortex system. For a loop lying in the $(x,y)$-plane, consider
the solution in, say, the $x=0$ plane.  The angular dependence of the
equation of motion is exactly the same as for the vortex-antivortex.
Since in the latter case the angular dependence could not be resolve,
the same is true for the string loop.  There are no bound state
solution.

To see this all more explicitly, let's look at the Dirac equation. To
describe bound states on loops of cosmic string we try the same
strategy as before and split the solution in a longitudinal and
transverse part
\be
\psi_{L,R} = \alpha(t,\vp) \beta_{L,R}(u,v) \xi_{L,R},
\ee
with $\xi_{L,R}$ eigenstates of the projection operator
\be
\sigma^0\sigma^\vp \xi_{L} = \pm \xi_{L}
\qquad
\sigma^0\sigma^\vp \xi_{R} = \mp \xi_{R}
\label{eigen}
\ee
Explicitly, the eigenspinors corresponding to the upper sign are
\be
\xi_{L} = 
\( \begin{matrix} 
\e^{-i\vp/2} \\
i \e^{i\vp/2}
\end{matrix} \),
\quad
\xi_{R} = 
\( \begin{matrix} 
-i\e^{-i\vp/2} \\
-\e^{i\vp/2}
\end{matrix} \),
\label{eigen1}
\ee
and the eigenspinors corresponding to the lower sign are
\be
\xi_{L} = 
\( \begin{matrix} 
\e^{-i\vp/2} \\
-i \e^{i\vp/2}
\end{matrix} \),
\quad
\xi_{R} = 
\( \begin{matrix} 
i\e^{-i\vp/2} \\
-\e^{i\vp/2}
\end{matrix} \).
\label{eigen2}
\ee
In contrast with the straight string, the eigenspinors are not
constant but depend on the angle $\vp$ along the string loop. The
$\vp$ dependence is needed to resolve the $\vp$-dependence in the
Dirac equation. For future use, we note that the spinors satisfy
$\sigma^r \xi_{L,R} = - \xi_{L,R}$, $\partial_\vp \xi_L = \pm \frac12
\xi_R$, and $\partial_\vp \xi_R = \mp \frac12 \xi_L$.

The longitudinal part of the Dirac equation is
$\(\partial_t - \sigma^\vp \frac{\partial_\vp}{h_\vp} \) \alpha
\xi_L = 0 $
%
and a similar equation for $\psi_R$.  Now since $\partial_\vp \xi_\pm
\neq 0$ an extra term appears and the equation reads
\be
\( \partial_t \alpha \mp \frac{\partial_\vp}{r}  \alpha \) \xi_L
+ \frac{1}{2r} \alpha \xi_R = 0
\label{long}
\ee
This cannot be solved as for the straight string.  First of all, the
equation depends through $h_\vp = r(u,v)$ on the longitudinal
coordinates, and the separation of variables is no longer valid.
Secondly, there is an extra term since the eigenvectors are
$\vp$-dependent, but one could hope that this term can be ``absorbed''
in the longitudinal part of the Dirac equation.

Nevertheless, let us try to split the frequency part and see where we
get. Try the Ansatz $\psi_{L,R} = \e^{-i E t}(\beta_1 \xi_+ + \beta_1
\xi_-)_{R,L}$, with $(\xi_\pm)_{L,R}$ the spinor eigenstates of
\eref{eigen}.  The transverse Dirac equation in cylindrical
coordinates is 
\bea
iE \beta_{1L}+
(D_r - i D_z + \frac{1}{2r}) \beta_{2L} 
+ i \lambda \phi \beta_{2R} &=& 0
\nonumber \\
iE \beta_{2L}+ (D_r + i D_z + \frac{1}{2r}) \beta_{1L} 
- i \lambda \phi \beta_{1R} &=& 0
\eea
and a similar equation for $L \leftrightarrow R$.  Now transform to
toroidal coordinates, see appendix \ref{s:coord} for details. Using
$D_r + i D_z = -(2i/a) \sin^2 \bar{\chi} (D_u+iD_v)$ and $1/(2r) =
\sin \chi \sin \bar{\chi}/(a \sinh v)$ we get
\bea
i \frac{a}{2} E \beta_{1L}+
\( \sin^2 {\xi} (D_u-iD_v) -
\frac {i\sin \xi \sin \bar{\xi}}{2\sinh v} \) \beta_{2L}
&+& \frac{a \lambda \phi}{2} \beta_{2R} = 0
\nonumber \\
i \frac{a}{2} E \beta_{2L}+
\( \sin^2 {\xi} (D_u+iD_v) -
\frac {i\sin \xi \sin \bar{\xi}}{2\sinh v} \) \beta_{1L}
&+& \frac{a \lambda \phi}{2} \beta_{1R} = 0
\label{Dloop}
\eea
If we follow the same strategy as for the straight string, we now want
to remove the $u$ dependence, i.e. the dependence on the coordinate
winding around the string.  This, however, seems impossible.  Remember
that $\chi$, $\bar{\chi}$ depend on $u$, so $\sin \chi$ contains a
$\e^{iu/2}$ and a $\e^{-iu/2}$ term.  The phase structure of the
equation above thus is
\be
\beta_{1L} + (\e^{iu} + 1 +\e^{-iu})\beta_{2L}
= \e^{inu}\beta_{2R}  
\ee
and similar for $1 \leftrightarrow 2$.  There is no way to remove the
phase dependence.  The situation is completely analogous to the vortex
antivortex system, whether $E$ is non-zero or not.

\section{Conclusions}

In this paper we have studied the fermionic spectrum on different
topological defects including systems of soliton-antisoliton systems.
We find that the number of zero modes is given by topological
arguments, just as in the case of a single soliton.  Any configuration
that can be continuously deformed to the trivial vacuum does not
support fermionic bound states.  Although this sounds like a trivial
conclusion, statements to the contrary can be found in the literature.  

To built up intuition, we first studied the simpler kink-antikink
systems. While a kink, antikink and kink-kink system all support one
fermionic zero mode, a kink-antikink has none due to the different
spinor structures of the fermionic mode associated with the kink and
antikink.  This is in complete agreement with topological
arguments. Moreover, we find that no massive fermionic states on any
system of kinks and anti-kinks exists.

Cosmic strings with winding number $n$ are known to support $n$
fermionic zero modes.  Accordingly, we find that a system of two
parallel BPS strings with $n=1$ each supports two fermionic zero
modes, while a string-antistring pair has no zero modes. In addition
we find that a loop of cosmic string (which can be approximated by a
string-antistring pair) has no fermionic zero modes. Note that this
contrasts with the results in \cite{Gadde,pr}.

Cosmic string loops stabilized by fermionic or bosonic currents,
so-called {\it vortons}, are believed to be important for
cosmology. Our results indicate that cosmic string loops with
fermionic currents on them do not exist.

Of course, we have neglected the back reaction of the fermions on the
bosonic fields and have only studied large loops. We believe, however,
that taking these two points into account will not change our main
conclusions.\\ \\ \\
\noindent
{\bf\large Acknowledgments} \\ We thank T.~Vachaspati and S.C.~Davis
for helpful discussions.  BH acknowledges S. Nicolis for bringing
reference (\cite{pr}) to her attention.  BH was supported by a CNRS
grant, MP acknowledges support from FOM Netherlands.\\ \\

\newpage
\appendix

\section{Cylindrical and toroidal coordinates}
\label{s:coord}

We use the flat space-time metric with signature $(+,-,-,-)$.  Consider the
three spatial dimensions with a metric of the form (\ref{3metric})
\be
\dd s^2 = - h_1^2 \dd x_1^2 - h_2^2 \dd x_2^2 - h_3^2 \dd x_1^2.
\ee
In Cartesian coordinates $(x_1,x_2,x_3) = (x,y,z)$ with $h_i=1$.
Cylindrical coordinates are most convenient to describe a straight
string aligned with the z-axis. Then $(x_1,x_2,x_3) = (r,\vp,z)$ and
$h_r = h_z=1, h_\vp=r$.  The relation to Cartesian coordinates is $(x,y,z) =
(r \cos \vp, r \sin \vp, z)$.  The Pauli matrices in cylindrical coordinates
are:
\be
\sigma^r=\( \begin{matrix}
0 & \e^{-i \vp} \\
\e^{i \vp} & 0
\end{matrix} \), \qquad
\sigma^\vp =\( \begin{matrix}
0 & -i \e^{-i \vp} \\
i \e^{i \vp} & 0
\end{matrix} \), \qquad
\sigma^z =\( \begin{matrix}
1 & 0 \\
0 & -1
\end{matrix} \).
\label{pauli_cyl}
\ee

Toroidal coordinates are most convenient to describe a loop of string
lying in the $x$-$y$-plane.  Then $(x_1,x_2,x_3) = (v,u,\vp)$ with
\bea
h_v = h_u &=&  \frac{\kappa}{\cosh v - \cos u},\nonumber \\ 
h_\vp &=& \frac{\kappa \sinh v}{\cosh v - \cos u}.
\label{h_tpr}
\eea
The relation with Cartesian coordinates is
\be
(x,y,z) = (\frac{\kappa \sinh v \cos \vp}{\cosh v - \cos u},\,
\frac{\kappa \sinh v \sin \vp}{\cosh v - \cos u} , \,
\frac{\kappa \sin u}{\cosh v - \cos u} ).
\label{coord_tor}
\ee
%
%
At the core of the string the radius of the loop is $\kappa$, and its arc
length $\kappa \dd \vp$ with $\vp \in [0,2\pi)$.  The coordinate $u \in
[0,2\pi)$ winds around the string core, whereas $v$ gives the radial
distance away from the core (but note, for $v\to \infty$ the string
core is approached, whereas $v\to 0$ corresponds to far away from the
core).  The Pauli matrices in toroidal coordinates read:
\bea
\sigma^v &=& \frac1{\cosh v - \cos u}
\( \begin{matrix}
\sin u \sinh v & \e^{-i \vp} (\cos u \cosh v -1) \\
\e^{i \vp} (\cos u \cosh v -1) & -\sin u \sinh v 
\end{matrix} \), \\
\sigma^u &=& \frac1{\cosh v - \cos u}
\( \begin{matrix}
-\cos u \cosh v+1 &  \e^{-i \vp}  \sin u \sinh v\\ 
\e^{i \vp} \sin u \sinh v & \cos u \cosh v -1  
\end{matrix} \), \\
\sigma^\vp &=&\( \begin{matrix}
0 & -i \e^{-i \vp} \\
i \e^{i \vp} & 0
\end{matrix} \).
\label{pauli_tor}
\eea
We also introduce the notation
\be
\chi = (u+ i v)/2.
\label{xi}
\ee
The equations of motion may be simplified using $\sin\chi \sin \bar{\chi} = \frac12
( \cosh v - \cos u) $ and $\sin^2\bar{\chi} = \frac12 ( 1 - \cos u
\cosh v -i \sin u \sinh v )$.

For 1-forms we use the notation 
\be
A = \sum A_i \dd x^i = \frac{A_i}{h_i} e^i
\label{1form}
\ee
with vierbein $e^i = h_i \dd x^i$.  For example, for a straight string in
cylindrical coordinates
\bea
A &=& A_\vp  \dd \vp = \frac{A_\vp}{r} e^\vp 
\nonumber \\
  &=& A_\vp \frac{\partial \vp}{\partial x} \dd x 
   + A_\vp \frac{\partial \vp}{\partial y} \dd y
  = - A_\vp \frac{y}{r^2} \dd x + A_\vp \frac{x}{r^2} \dd y
  = A_x \dd x + A_y \dd y.
\eea
And similarly for higher forms: $F = F_{12} \dd x^1 \dd x^2 =
\frac{F_{12}}{h_1 h_2} e^1 e^2$.

\subsection{Asymptotics}

The core of the straight string aligned with the z-axis is approached
in the limit $r = \sqrt{x^2+ y^2} \to 0$ (and $\vp$ ill defined),
whereas asymptotically far corresponds to $r \to \infty$.

In toroidal coordinates the core of the string is approached in the
limit $v \to \infty$ (and $u$ ill defined).  Then $(x,y,z) \to (\kappa
\cos \vp, \kappa \sin \vp,\kappa \sin u \e^{-v} )$ and
$(h_v,h_u,h_\vp) \to (\kappa \e^{-v},\kappa \e^{-v}, \kappa)$.  The
core of the string loop is at $r = \sqrt{x^2+y^2} = \kappa$.  Write $r
= \kappa + \delta r$, then we can express the limit $v \to \infty$ in
terms of $\delta r \to 0$.  More precisely
\be
v = -{\rm arccosh} \( \frac{r^2 + \kappa^2}{r^2 -\kappa^2} \) 
\;\stackrel{\delta r \to 0} {\longrightarrow}\;
\log \( \frac{\kappa} {\delta r} \),
\label{core_limit1}
\ee
and thus
\be
\lim_{v\to \infty} e^{-v} \sim  \lim_{\delta r \to 0} \frac{\delta r}{\kappa}
\label{core_limit2}
\ee
($r,\kappa$ have dimension of length, while $u,v,\vp$ are dimensionless.)\\

Spatial infinity corresponds to $u = v \to 0$. In this limit $\cosh v
- \cos u \sim v^2$, $\sinh v \to v$ and thus
\bea
(x,y,z)
&\stackrel{u=v \to 0} {\longrightarrow}&
(\frac \kappa v \cos \vp, \frac{\kappa}{v} \sin \vp, \frac \kappa v) 
\nonumber \\
(h_v,h_u,h_\vp)  
&\stackrel{u=v \to 0} {\longrightarrow}&
(\frac{\kappa}{v^2}, \frac{\kappa}{v^2}, \frac{\kappa}{v})
\label{inf_limit1}
\eea
and 
\be
\lim_{u=v\to 0} v \propto  \lim_{R\to \infty} \frac \kappa R  
\propto  \lim_{r\to \infty} \frac \kappa r
\label{inf_limit2}
\ee
with $R = \sqrt{x^2+y^2+z^2} = \sqrt{r^2+z^2} \propto r$. In the limit
$\kappa \ll R$, the distance from the string core approaches the distance
from the origin $R$.

Finally, the limit $u = \pi, v\to 0$ corresponds to the origin, the
center of the loop.  Indeed, in this limit
\bea
(x,y,z) 
&\stackrel{u \to \pi, v \to 0} {\longrightarrow}&
(\frac \kappa2 v \cos \vp, \frac \kappa2 v \sin \vp, 0) 
\nonumber \\
(h_v,h_u,h_\vp) 
&\stackrel{u\to \pi,v \to 0} {\longrightarrow}&
(\frac \kappa2,\frac \kappa2 ,\frac \kappa2 v) 
\label{origin_limit1}
\eea
and 
\be
\lim_{u\to\pi,v\to 0} v \propto  \lim_{r\to0} \frac r\kappa.  
\label{origin_limit2}
\ee

\subsection{Notation and convention}
\label{s:conv}

Most of the solutions are given in terms of eigenspinors, and are independent of the
basis of $\gamma$-matrices.  When we give explicit solutions, we
use the chiral basis:
\be
\gamma^\mu = \( \begin{matrix} 
0 & \sigma^\mu \\
\bar{\sigma}^{\mu} & 0 \\
\end{matrix} \)
\ee
with $\sigma^\mu = (1,\sigma^i)$, $\bar{\sigma}^\mu = (1,-\sigma^i)$,
and $\sigma^i$ the Pauli matrices.

\section{Asymptotic solutions of \cite{Gadde}}
\label{s:gadde}

In this appendix we show explicitly that there are no fermionic zero
modes on a circular loop of cosmic string.  This is in contrast with
the results in \cite{Gadde}. Following \cite{Gadde} we remove half of
degrees of freedom by considering Majorana fermions with $\psi_R =
\psi_L^c = i\sigma \psi_L^*$.  The equation to solve is then
\eref{Dloop} with $E=0$.  Using the Ansatz
\be \beta = X \e^{i l (\chi + \bar{\chi})} + Y^* \e^{i (n-l) (\chi +
\bar{\chi})}
\label{ansatz}
\ee
this gives the two coupled differential equations
\bea
\(  
\sin^2 \bar{\chi} \(\partial_{\bar{\chi}} + i(l-\frac{n}{2} a )\)
+ \frac {i\sin \chi \sin \bar{\chi}}{2\sinh v} 
\) X
&=&  \frac{\kappa m f}{2} Y
\nonumber \\
\(  
\sin^2 {\chi} \(\partial_{\chi} - i(n-l-\frac{n}{2} a )\)
- \frac {i\sin \chi \sin \bar{\chi}}{2\sinh v} 
\) Y
&=& \frac{\kappa m f}{2} X
\label{XY}
\eea
with $m=\lambda \eta$ the vacuum fermion mass. These equations agree
with those found in \cite{Gadde}.

\paragraph{Infinity.}

Let us first consider the behavior at infinity and write
\be
\chi = \frac12 (u+ i v) = \frac1t \e^{i \vp}.
\label{t}
\ee
Spatial infinity is approached in the limit $u = v \to 0$, or
equivalently $t \to \infty, \vp \to \pi/4$.  Further, the gauge
field and Higgs field approach their vacuum values and thus $a(v)
\to 1$ and $f(v)\to 1$.  Then using the approximation $\sin \chi \approx \chi$
and after redefining the fields
\be
X = A \chi \sqrt{\frac{2i\chi \bar{\chi}}{\chi -\bar{\chi}}},
\qquad
Y = B \bar{\chi} \sqrt{\frac{ 2i\chi \bar{\chi}}{\chi -\bar{\chi}}},
\ee
(\ref{XY}) becomes
\bea
\chi \bar{\chi} \( \partial_{\bar{\chi}} - i p \) A &=& \frac{\kappa m}{2} B
\nonumber \\
\chi \bar{\chi} \( \partial_{\chi} - i p \) B &=& \frac{\kappa m}{2} A
\eea
where we defined $p = n/2-l$.  Combining the two equations gives
\be
\chi \bar{\chi} \( \partial_{\chi} - i p \)
\chi \bar{\chi} \( \partial_{\bar{\chi}} - i p \) A
= \(\frac{\kappa m}{2}\)^2 A.
\label{Ainf}
\ee
Using (\ref{t}) we write this in terms of $t,\vp$ to take the
limit $r \to \infty$.  The partial derivatives become
\be
\partial_\chi = -\frac{t^2 \e^{-i\vp}}2 
(\partial_t + \frac{i}t \partial_\vp),
\qquad
\partial_{\bar{\chi}} = -\frac{t^2 \e^{i\vp}}2 
(\partial_t - \frac{i}t \partial_\vp)
\ee
and $\chi \bar{\chi} = t^{-2}$.  Using these expressions in (\ref{Ainf})
and keeping only the terms with highest power in $t$ gives:
\be
\frac14 \partial_t^2 A - \frac{i}{4t} \partial_t \partial_\vp A - 
\(\frac{m\kappa}{2}\)^2 A + \mcO(t^{-2}) = 0
\ee
from which it follows that
\be
\lim_{r\to \infty} A = C \e^{-m\kappa t} 
\propto \e^{\pm m r}
\ee
with $C$ a normalization constant, and where we used $t \propto v^{-1}
\propto \kappa/r$.  There is a similar solution for $B = \e^{-i\pi/4}
A$. Putting back all the factors we get for the normalizable solution:
\be
\beta = \frac{C \e^{i \pi/4} \e^{-m\kappa t}}{t^{3/2}}
\(\e^{i l \sqrt{2}/t} + \e^{i \pi/4} \e^{-i (n-l) \sqrt{2}/t} \).
\ee

\paragraph{Loop core.}

The core of the loop is approached in the limit $v\to \infty$ or $\chi
\to iv/2 \to i \infty$. In this limit $\sin \chi \to i \e^{v/2}$.
Further $f , a \to 0$ at the core.  Taking this limit in the
Weyl equation (\ref{XY}), only the first term survives and gives
\bea
\(\partial_{\bar{\chi}} + i l \) X = 0 
\qquad &\Rightarrow& \qquad
X \sim f_X(\chi) \e^{- i l \bar{\chi}}  
\nonumber \\
\(\partial_\chi - (n-l) \) Y = 0 
\qquad &\Rightarrow& \qquad
Y \sim f_Y(\bar\chi) \e^{ i (n-l) \chi} 
\eea
Then
\bea
\beta &=& f_X(\chi) \e^{i l \chi} + f_Y(\chi) \e^{i(n-l)\chi}
\nonumber\\
 &=& f_X(v) \e^{-\frac{lv}{2}} +f_Y(v) \e^{-\frac{(n-l)v}{2}} 
\ \ \ {\rm for} \ v\to\infty
\eea
Since the functions $f_X(v), f_Y(v)$ are unconstrained, nothing much
can be said about the number of solutions at the string core.

\paragraph{Origin.}

Finally consider the zero mode solution at the origin $(u,v) \to
(\pi,0)$ or $(\chi,\bar{\chi}) \to (\pi/2,\pi/2)$, where $f\rightarrow 1$, $a
\rightarrow 1$. Note that this case was not
studied in \cite{Gadde}. The Weyl
equation (\ref{XY}) becomes
\bea
\( (\partial_{\bar{\chi}} -i p) - \frac{1}{2 (\chi -\bar{\chi})} \) X
&=& \frac{\kappa m}{2} Y
\nonumber \\
\( (\partial_{{\chi}} -i p)+ \frac{1}{2(\chi -\bar{\chi})} \) Y
&=& \frac{\kappa m}{2} X
\eea
where we have used the abbreviation $p=\frac{n}{2} - l$. 

The problem is already apparent here: $\chi-\bar \chi \to 0$ and the second term
on the left hand side of both equations blows up. Redefining the fields
\be
X = A \sqrt{\frac{2i}{\chi - \bar{\chi}}},
\qquad \qquad
Y =  B \sqrt{\frac{2i}{\chi - \bar{\chi}}}
\ee
the equations simplify to
\bea
(\partial_{\bar{\chi}} -ip) A &=& \frac{\kappa m}{2} B
\nonumber \\
(\partial_{\chi} -ip) B &=& \frac{\kappa m}{2} A
\eea
which can be combined into a single equation for $A$
\be
(\partial_\chi -ip) (\partial_{\bar{\chi}} -ip) A = \(\frac{\kappa m}{2}\)^2 A
\ee
and likewise for $B$.  Explicitly, we have:
\be
\partial_\chi \partial_{\bar{\chi}} A 
-ip(\partial_\chi + \partial_{\bar{\chi}})A
-\(p^2 + (\kappa m/2)^2 \) A =0.
\ee
We will try two type of Ans\"atze.  First, we choose $A = A(\chi +
\bar{\chi})= A(u)$, which implies $A''=2ipA'+(p^2+(\kappa m/2)^2)A$
and the prime denotes the derivative with respect to $u$. This has as
solution
\be
A = C \e^{(\pm\kappa m/2+ip)u} 
\quad \stackrel{u \to \pi,v\to0}{\longrightarrow} \quad
{\rm const.}
\ee
Further $B \propto A$ where the constant of proportionality is given by 
the equations.  

The second type of Ansatz is $A=A(-i (\chi-\bar\chi))=A(v)$, which
gives $A''=(p^2+(\kappa m/2)^2 A$ where the prime denotes derivative
with respect to $v$. This has as solution
\be A = C \e^{\pm\sqrt{p^2+(\kappa m /2)^2}v} \quad \stackrel{u \to
\pi,v\to0}{\longrightarrow} \quad {\rm const} \ee Both solutions do
not tend to zero at the origin.  On the contrary, plugging the above
expressions back in, we find that zero mode solution
\be
\beta = \sqrt{\frac{2}{v}}
\(A \e^{i l u} + B^* \e^{i(n-l) u}\)
\quad \stackrel{u \to \pi,v\to0}{\longrightarrow} \propto 1/v
\ee
Although the solution blows up in the origin as $1/v$ the integral
$\int \dd v \dd u \sqrt{g_2} \psi^\dagger \psi$ is finite.
Nevertheless, this is not a valid solution.  The spinors
(\ref{eigen1}), (\ref{eigen2}) are only well defined at the origin if
$\beta \to 0$. Note that a similar argument holds for the full
$z$-axis with $v=0$ and $u\neq 0$. The solution is thus singular on
the symmetry-axis.  Non-zero energy will affect the exponents in
$A,B$, but not the dominant behavior of $\beta$ in the limit $u\to
\pi, v\to 0$.  And thus non-zero energy cannot alleviate this problem.

 \end{document}